# What is a quantum-mechanical "weak value" the value of?

**Bengt E Y Svensson**

**Abstract** A so called "weak value" of an observable in quantum mechanics (QM) may be obtained in a weak measurement + post-selection procedure on the QM system under study. Applied to number operators, it has been invoked in revisiting some QM paradoxes (*e.g.,* the so called Three-Box paradox and Hardy's paradox). This requires the weak value to be interpreted as a *bona fide* property of the system considered, *a par* with entities like operator mean values and eigenvalues. I question such an interpretation; it has no support in the basic axioms of quantum mechanics and it leads to unreasonable results in concrete situations.

**Keywords** Weak value, number operator, Hardy's paradox, interpretation of QM

B.E.Y. Svensson
Theoretical High Energy Physics, Department of Astronomy and Theoretical Physics, Lund University,
Sölvegatan 14A, SE-223 62 Lund, Sweden
e-mail: Bengt_E_Y.Svensson@thep.lu.se

## 1 Introduction

The pioneering work by Aharonov, Albert and Vaidman [1] introduced the concept of 'weak measurements' of an observable in quantum mechanics (QM) in combination with the procedure of 'post-selection'. The result of such a weak measurement + post-selection can be expressed in terms of the 'weak value' $({}_f\hat{O}_{in})_w$ of the observable $O$ under study. For an initial ('pre-selected') state $|in>$ and a final ('post-selected') state $|f>$, this weak value is defined by [1]

$$({}_f\hat{O}_{in})_w := \frac{<f|\hat{O}|in>}{<f|in>} \qquad (1)$$

The basic idea of weak measurement + post-selection has since attracted much interest; for recent reviews with further references see, *e.g.*, [2, 3, 4]. Not the least has the field opened up new tools for experimentalists to investigate aspects of phenomena that were thought impossible earlier. These include determining (even if only statistically) the trajectories in a double slit experiment without destroying the interference pattern [5], and directly measuring the wave function [6]. The use of the technique for amplification purposes has been spectacular [7-9]. Weak measurements have also been employed to illuminate the fundamental difference between classical and quantum mechanics exhibited by violation of Leggett-Garg inequalities [3, 10–14], also called "Bell inequalities in time".

In addition, weak values of number operators have been invoked to revisit what are conceived as QM paradoxes, like the so called Three-Box Paradox [15-18] and Hardy's Paradox [19–22]. It is even claimed that some of the so called 'counter-factual' statements in these 'paradoxes' can be made 'factual' by using weak measurements, quantified in terms of weak values, instead of the ordinary 'strong' measurements. These claims rely on an straight-forward interpretation of a weak value on *a par* with, *e.g.*, a usual mean value. That is, one ascribes a meaning to a weak value as an 'ordinary' property – like a probability, the number of particles, *etc.* – of the system under investigation.

In this paper, I investigate these very basic interpretational issues. In particular, I criticize such a realistic, straightforward interpretation (for short "RSFI" in the sequel). My emphasis is on a very fundamental (and also elementary) level: given that a weak value $({}_f\hat{O}_{in})_w$ has the definition and the operational connotations that it has, are there grounds for seeing it as a *bona fide,* 'ordinary' property (like a mean value $<\hat{O}>$; see section 2 for a more explicit description of what I mean by an 'ordinary' property )? I approach this problem from different angles: Do the conventional, basic rules – the 'axioms' – of QM have any bearing? Is an RSFI consistent with varying the entities defining $({}_f\hat{O}_{in})_w$ , in particular with varying the post-selected state $|f>$?

---

[1] It can be measured in a so called indirect measurement scheme in which the examined 'system' starts in the initial state $|in>$ ('pre-selection'), interacts weakly with a suitable 'meter', and is then projected into a final state $|f>$ ('post-selection'). Following this procedure, the weak value can then be obtained by suitably reading off the meter. See, *e.g.*,[2, 3,4] for further details.

Note that I do not here question the weak value as a measureable entity pertaining to the system under study, nor the many ingenious ways experimentalist have exploited the weak measurement + post-selection idea ([5 – 9, 11 – 14]; see also [3] for a brief review). What I do question is whether a weak value can be given an 'ordinary' meaning.

My focus is on the use of weak values for number operators. However, I begin in section 2 with some more general observations and remarks on weak values. In sections 3 – 5, I then study how weak values of number operators are used in connection with (variants of) the Three-Box Paradox, Hardy's Paradox and a Mach-Zehnder interferometer, and show that an RSFI of these weak values leads to results which seem to lack a reasonable meaning. An appendix is devoted to the relation between weak values of number operators and the so called Aharonov-Bergmann-Lebowitz (ABL) probability formula [23].

## 2  Interpreting a weak value – some general remarks

In the application of weak values, it is often taken for granted, explicitly or implicitly, that en entity like $(_f \hat{O}_{in})_w$ has the same basic meaning as an average value $< \hat{O} >$. This is done by Aharonov and collaborators when they treat the Three-box Paradox and the Hardy's Paradox as I will describe in some detail in sections 3 and 4 below. It is also what they do when they discuss "negative kinetic energy" in tunneling experiments [15]. And Vaidman [24, 25] has argued that this is legitimate:

> *If we are certain that a procedure for measuring a certain variable will lead to a definite shift of the unchanged probability distribution of the pointer, then there is an element of reality: the variable equal to this shift.* [24]

Sometimes they or other protagonist of the weak-measurement + post-selection procedure even state that there are experimental support for ascribing such an 'ordinary' meaning to the weak value [2, 25–28].

Ever since the path-breaking paper [1] by Aharonov *et al* there has been much discussion of what a 'weak measurement' really can accomplish and what significance one can and should ascribe to a 'weak value'; the whole discussion can be retrieved from [2, 15] and references therein. Of the previous criticisms of weak values that seems most similar to mine, I should mention [29, 30] (to which Aharonov and Vaidman replied [31]) and [32], of the more recent ones [33] and, even more [34], in which Kastner explicitly points out that a weak value "should be thought of as an amplitude and not an expectation value at all.". Indeed, the weak value is the ratio of two amplitudes. My arguments may be seen as a further deepening of Kastner's criticism.

It is without question that the weak value $(_f \hat{O}_{in})_w$ is an "element of reality" in the general sense described by Vaidman [24] in the quotation above: the weak value is an experimentally accessible entity. And part of the ingenuity of introducing the weak measurement + post-selection procedure is that it allows direct measurement of such a ratio of two amplitudes. But the very fact that an entity can be measured does in no way entails its interpretation. Experiments really have little bearing on the *interpretation* of the weak value; it must be the theory that decides what *meaning* to ascribe to it. Nor can the fact that experiments confirm some of the predictions for weak values be taken as an argument for the meaning of the weak value. For examples, experiments on Hardy's Paradox [21, 22] are indeed ingenious in their own right. However, what they find is but a consequence of QM: they do confirm the results of Aharonov *et al*, results that are entirely based on QM. To put it in another way: should the experimentalists have found disagreement with the results derived by Aharonov *et al*, they would have found a violation of QM.

In conclusion, the facts that $(_f \hat{O}_{in})_w$ can be measured and that the results in concrete experimental situations agree with theoretical predictions cannot be taken as support for an RSFI of $(_f \hat{O}_{in})_w$ as an 'ordinary' property.

In searching for an argument that could motive what meaning to ascribe to $(_f \hat{O}_{in})_w$, being the ratio of two amplitudes, it is natural to go back to the very foundations of QM, to the basic rules that govern all the endeavor: QM (for a finite system) lives in a (separable) Hilbert space $\mathcal{H}$, an observable $O$ is represented by a self-adjoint operator $\hat{O}$ in $\mathcal{H}$, the result of (strongly) measuring the observable $O$ is an eigenvalue $o_n$ of $\hat{O}$, a state is describes by a density matrix $\rho$ in $\mathcal{H}$, and the probability of obtaining a particular eigenvalue $o_n$ equals the trace $Tr\,(\hat{P}_n\,\rho)$ with $\hat{P}_n$ the projector onto the subspace of $\mathcal{H}$ spanned by the eigenvectors of $\hat{O}$ with eigenvalue $o_n$. These rules imply, *i.a.*, that one may ascribe experimental and conceptual meaning to theoretical entities like the eigenvalues of $\hat{O}$, the probability of obtaining an eigenvalue, and the mean value of the operator in a given state. These are examples of what I call the 'ordinary' meaning of the theoretical concepts of QM.

*Nowhere in these basic rules can I find any motivation for interpreting a ratio of two amplitudes, like the weak value $(_f \hat{O}_{in})_w$, as an 'ordinary' property in this sense.*

But the basic rules are not sacred [2]. No-one would object if they were supplemented with further insight provided this insight were in agreement with all previous experience and with other 'reasonable' arguments, in particular also being compatible with the usual meaning of the concepts involved. To try to find out whether such an approach to the meaning of the weak value $(_f \hat{O}_{in})_w$ is a practical way forward, I will devote the following sections to investigating some concrete applications of weak values to the cases in which the operator $\hat{O}$ is a number operator.

---

[2] For example, some authors [35,36] have proposed to consider (essentially) the weak value as an extended probability, which might even take complex values. I shall not follow that track here; it is not what I call "compatible with the usual meaning of the concepts involved".

## 3 The Three-Box Paradox

One (thought) experiment which Aharonov *et al* analyzed in terms of weak values is the Three-Box Paradox. Imagine, they say [15 − 17], a single QM particle in any one of three boxes *A*, *B* and *C*. The QM states representing the particle in box *A* is denoted $|A>$, with corresponding notations for *B* and *C*. Let the particle be described by a superposition given by the initial ('pre-selected') state $|in> = (|A> + |B> + |C>)/\sqrt{3}$. Suppose further that the particle is later found in the ('post-selected') state $|f> = (|A> + |B> - |C>)/\sqrt{3}$. Moreover, consider an (intermediate in time) measurement of the projection operator $\hat{A} := |A><A|$. It is a number operator which, upon measurement, counts the number of particles in the box *A*; its mean value gives the probability of finding the particle in box *A*. In the same way, $\hat{B} := |B><B|$ is the number operator for the particle in box *B*, and similarly for *C*. One is interested in the probability $prob_A(\text{in } A)$ for finding the particle in box *A* when measuring $\hat{A}$, as well as the corresponding probabilities for *B* and *C*.

Consider first an ordinary, 'strong' measurement of the respective projection/number operator. The ABL rule as presented in the Appendix applies and gives, for the pre- and post-selected state as given above,

$$prob_A(\text{in } A) = 1 = prob_B(\text{in } B) \quad \text{while} \quad prob_C(\text{in } C) = 1/5 \qquad (2)$$

At first, there seems to be a paradox here. Not only does the total probability to find the particle in any box exceed 1, it is with certainty – or at least with probability 1 – to be found both in box *A* and in box *B*. But the paradox disappears when one realizes that the results apply to *different* projective measurements, which certainly cannot be performed in conjunction without each measurement heavily disturbing ('collapsing') the system and thereby creating totally new conditions for a subsequent measurement. In other words, the paradox only appears if one allows 'counter-factual' statements which require, for their verification, measurements that are non-implementable in an ordinary, 'strong' measurement scheme.

Could weak measurements come to the rescue? It is certainly possible to implement them. Indeed, nothing forbids one to do all three weak measurements of the number operators $\hat{A}$, $\hat{B}$ and $\hat{C}$ successively on the given pre- and post-selected states: the fact that they are weak measurements ensures that one may (approximately) disregard measurement disturbances. Thus, with weak measurements, there are not necessarily any counter-factual statements involved.

The relevant weak values are

$$(_f\hat{A}_{in})_w = 1 = (_f\hat{B}_{in})_w. \qquad (3)$$

On the other hand,

$$(_f\hat{A}_{in})_w + (_f\hat{B}_{in})_w + (_f\hat{C}_{in})_w = [_f(\hat{A} + \hat{B} + \hat{C})_{in}]_w = 1, \qquad (4)$$

which together imply

$$(_f\hat{C}_{in})_w = -1 \ ! \tag{5}$$

(Of course, this could also be calculated directly from the explicit expression for $(_f\hat{C}_{in})_w$.) Consequently, with an 'ordinary' meaning of these weak values in which one interprets the weak value as a *bona fide* value of a number operator, one arrives at the mind-boggling result that there is minus one particle in box *C*!

Let me see how this stands further scrutiny.

Let me first note that the result for the weak values of $\hat{A}$ and $\hat{B}$ being unity is in agreement with the strong, ABL-values in eq. (2) being unity (see the Appendix below).

The 'strong' values of a projection operator are its eigenvalues, 1 and 0. It is this fact that, from the basic rules of QM reviewed in section 2 above, legitimates the result of a measurement of a projection operator to be interpreted as the (relative) number of particles in the respective box, and its mean value as the probability of finding the particle in that box. But how legitimate is it to interpret the *weak value* of a projection operator here as 'the', or even 'a', particle number, or as a probability?

An important difference to an ordinary mean value is that a weak value depends not only on the pre-selected state, $|in>$, but also on the post-selected one, $|f>$. In other words, whatever it is, it is not only referring to the system in the initial, pre-selected state, but to the whole set-up of the situation being analyzed. And, by considering different combinations of the basis states $|A>$, $|B>$ and $|C>$ for $|f>$, one may get essentially any result for the weak value of a number operator. In fact, one may look upon the post-selected state as a kind of filter which can be tuned as one wants. Expressed differently, even for a given pre-selected state $|in>$, like the one chosen here, there is still the freedom of choosing the post-selected state $|f>$ in any way one likes. The particular one $|f> = (|A> + |B> - |C>)/\sqrt{3}$ used in the formulation above has no preordained physical meaning. It is just one among a multiple continuum of choices, none of which seems more natural than the other. I devote a section in my review [3] to a more extensive study of this ambiguity.

Attempts have also been made by Aharonov and collaborators to give, in terms of thought experiments, physical meaning to, *e.g.*, the value $-1$ for the weak value of a number operator. However, as I show in [3] for the particular case treated by Rohrlich and Aharonov in [37], their explanation, too, is heavily dependent on the filtering function of post-selection. Therefore, such a thought experiment carries no further explanatory power.

In conclusion, in this case of the so called Three-Box Paradox, there are strong arguments disfavoring an RSFI of the weak values as 'ordinary's values of number operators. Besides not having any motivation from the basic postulates of QM, the weak values here can take essentially any complex value, not in any obvious way related to a number. More conceptually, for a given initial state, the weak values depend also on the choice of the final, post-selected state and cannot be interpreted merely as a property of the initial state alone: a weak value like $(_f\hat{C}_{in})_w$ cannot simply tell how many particles there are in box C.

## 4 Hardy's Paradox.

In [19], Hardy outlined a QM experiment that implies a paradox if interpreted in classical terms. Aharonov and collaborators gave an analysis of the experiment [20] in terms of weak values, in which they again relied on an RSFI of weak values of number operators. In this section, I discuss the legitimacy of this interpretation. My treatment here relies heavily on the paper [20] by Aharonov *et al*.

Two experimental groups have performed experiments [21, 22] and confirmed the results of Aharonov *et al*. As I made clear in section 2 above, this has no bearing on the *interpretation* of the weak values.

The setting for Hardy's Paradox is schematically presented in figure 1. It is assumed that the beam-splitters are ideal 50-50 splitters, that the arms of the Mach-Zehnder interferometers for the electron respectively for the positron are of equal length, and that there are no other obstructions in the arms but the annihilation region in the interacting arms $I_p$ and $I_e$ where annihilation is assumed to occur with unit probability. Also, the electron and positron are assumed to enter the apparatus simultaneously. The assumption of simultaneous passage of a positron-electron pair through the whole set-up should therefore be valid.

The analysis proceeds by successively considering what happens at the relevant moment of the particles' traversal through the apparatus. In a hopefully self-explanatory notation, the transition at the beam-splitter $BS_p1$ is described by

$$|p> \xrightarrow{BS_p1} = \{|N_p> + i|I_p>\}/\sqrt{2}, \qquad (6)$$

with similar transitions at the other beam-splitters. The state of the electron-positron pair just before it enters the second pair of beam-splitters, *i.e.* after the possible annihilation in the $I_e$-$I_p$-intersection, is without an $|I_p> \otimes |I_e>$ - term, and reads

$$|in> := \{|N_p> \otimes |N_e> + i|I_p> \otimes |N_e> + |N_p> \otimes i|I_e>\}/\sqrt{3}. \qquad (7)$$

At the second pair of beam-splitters this turns into

$$\{-|D_p> \otimes |D_e> + i|D_p> \otimes |B_e> + i|B_p> \otimes |D_e> - 3|B_p> \otimes |B_e>\}/\sqrt{12}. \qquad (8)$$

The paradox is now the following. In the conventional interpretation, and noting that the state (8) can be written

$$\{(-|D_p> + i|B_p>) \otimes |D_e> + i |D_p> \otimes |B_e> - 3 |B_p> \otimes |B_e> \}/\sqrt{12} =$$

$$= \{ - i \sqrt{2}\ |I_p> \otimes |D_e> + i |D_p> \otimes |B_e> - 3 |B_p> \otimes |B_e> \}/\sqrt{12}\ , \qquad (9)$$

one sees that a click in the detector arm $D_e$ means that the positron must have traveled through the arm $I_p$, and similarly for $D_p$ with respect to $I_e$. Therefore, from a simultaneous click in $D_e$ and $D_p$ it seems reasonable to conclude that both particles should have gone through their respective interacting arm $I_p$ and $I_e$, in which case they should have annihilated and therefore not have been able to reach any detector. This is Hardy's paradox.

Aharonov et al [20] now want to check this by actually measuring through which arms the particles went. Of course, this cannot be done by an ordinary, 'strong' measurement, since that would 'collapse' the wave-function and thus destroy coherence. Instead, they invoke weak measurements of number operators, i.e., single particle occupation operators like $|N_p><N_p| =: \hat{N}_p$ and pair occupation operators like $\hat{N}_p \otimes \hat{I}_e$, etc. They express their results in terms of the weak values of these operators. The pre-selected state they choose is the state $|in>$ of eq. (7), and the post-selected state is

$$|f> \ = |D_p> \otimes\ |D_e>, \qquad (10)$$

i.e., simultaneous clicks in the detectors $D_p$ and $D_e$.

They deduce

$$[_f(\hat{I}_p \otimes \hat{I}_e)_{in}]_w = 0 \qquad (11)$$

i.e., vanishing weak value for the simultaneous appearance of the particles in the interacting arms, which is not unreasonable. Moreover

$$[_f(\hat{N}_p \otimes \hat{I}_e)_{in}]_w = 1 = [_f(\hat{N}_e \otimes \hat{I}_p)_{in}]_w\ , \qquad (12)$$

which Aharonov et al interpret as implying that there are *two* particle pairs in the apparatus simultaneously: one pair in the $N_p$- and $I_e$-arms, the other in the $N_e$- and $I_p$-arms. However, "(q-)uantum mechanics solves the paradox in a remarkable way" by giving

$$[_f(\hat{N}_p \otimes \hat{N}_e)_{in}]_w = -1, \qquad (13)$$

"*i.e.* that there is also *minus* one electron-positron pair in the non-overlapping arms which brings the total down to a single pair." (Quotes are from [20]).

As is evident, the arguments by Aharonov et al rely heavily on an 'ordinary' interpretation of weak values: the weak values of the pair occupation operators are straight-forwardly, and without further ado, identified with the number of pairs in the respective arms of the interferometer.

I question whether this is legitimate. Again, I base this doubt partly on the fact that another choice of the final state $|f>$ than that of eq. (10) would have given other values for the weak values.

True, the constraints on the choice of the final, post-selected state in this case seems more well-motivated than in the Three-Box Paradox treated above: Given the set-up and the questions asked, it is very natural to choose $|f> = |D_p> \otimes |D_e>$. But nothing fundamental would forbid one to choose other final states also in this case, even if these states might seem somewhat contrived. To illustrate this, I will, however, avoid the slightly more formal complications of the double Mach-Zehnder interferometer of the Hardy set-up and, in the next section, apply my arguments to a single Mach-Zehnder interferometer.

## 5  A single, slightly generalized Mach-Zehnder device.

Consider then the simple Mach-Zehnder interferometer as illustrated in figure 2, but with a more general second beam-splitter BS2. In fact, the most general transformation at a beam-splitter obeying unitarity (= probability conservation) and time-reversal invariance, can be written in obvious matrix notation (and with the same notational conventions for the states as above)

$$\begin{pmatrix} |I> \\ |N> \end{pmatrix} = \begin{pmatrix} q & i\,r\,exp(i\beta) \\ i\,r\,exp(-i\beta) & q \end{pmatrix} \begin{pmatrix} |B> \\ |D> \end{pmatrix}, \qquad (14)$$

where q, r and β are real numbers and $q^2 + r^2 = 1$. The successive transitions will then be (the first beam splitter is assumed to be a perfect 50-50 splitter)

$$|p> \xrightarrow{BS1} \tfrac{1}{\sqrt{2}}\{|N> + i|I>\} \xrightarrow{BS2}$$

$$\xrightarrow{BS2} \tfrac{i}{\sqrt{2}}\{(q + r\,exp(-i\beta))\}|B> + \tfrac{1}{\sqrt{2}}\{(q - r\,exp(i\beta))\}|D>\}. \qquad (15)$$

Now, let the state just before the particle enters the second beam-splitter be the pre-selected state, *i.e.*, choose

$$|in> = \tfrac{1}{\sqrt{2}}\{|N> + i|I>\}, \qquad (16)$$

and one or the other of $|B>$ and $|D>$ as the post-selected state $|f>$. The expressions for the weak value of the respective number operators then read

$$[_{f=D}(\hat{N})_{in}]_w = \frac{q}{q - r\exp(i\beta)} \qquad (17)$$

and

$$[_{f=B}(\hat{N})_{in}]_w = \frac{r}{r + q\exp(i\beta)}, \tag{18}$$

with similar results for the weak values of $\hat{I}$.

To say the least, these expressions are difficult to interpret as realistic particle numbers or as ordinary probabilities. Not only would, in such an interpretation, the number of particles in the arms depend on how they are detected, *i.e.*, on the choice of the state $|f>$. Also, the 'number' could take any value, *e.g.*, $[_{f=D}(\hat{N})_{in}]_w = -1$ for $q = 1/\sqrt{5}$, $r = 2/\sqrt{5}$, $\beta = 0$. I think no-one would bet on having minus one particle in the *N*-arm!

One might argue that one should consider, instead, the real part of these expressions since it is the real part of a weak value that, in the measurement scheme (see footnote 1), is most directly related to the pointer position of the meter. But this does not help. Indeed, one finds, *e.g.*,

$$Re\,[_{f=D}(\hat{N})_{in}]_w = \frac{q(q - r\cos\beta)}{1 - 2qr\cos\beta}, \tag{19}$$

which again seem to defy a reasonable interpretation.

These facts cast severe doubts on the whole enterprise of interpreting a weak value in a RSFI fashion as anything reflecting a property of the pre-selected state *per se*.

As a side-remark, it might be interesting to evaluate the corresponding ABL probabilities (see the Appendix below). For example

$$prob\,(\hat{N} = 1\,|\,|f> = |D>, |in>) = r^2, \tag{20}$$

*i.e.*, the probability of finding the particle in the *N*-arm, given that it ends up in the *D*-detector is given by the overlap $|<D|N>|^2$ which is not an unreasonable result.

## 6 Summary

This paper makes two main points:

(1) The weak value is the ratio between two amplitudes. In section 2, I argue that neither experiments nor the basic rules of quantum mechanics can be invoked to motivate a realistic, straight-forward interpretation (an "RSFI") of such a construct.

(2) Moreover, in concrete cases – the Three-Box paradox (section 3), Hardy's paradox (section 4), a slightly generalized single Mach-Zehnder set-up (section 5) – it seems difficult to uphold an RSFI for the weak values of the number operators involved. The main reasons are twofold: the weak value is not a property of the initial state only, but depends also on the choice of the final ('post-selected') state, and non-sensible (complex) values for what is supposed to be a number may easily be obtained.

This casts doubts on any sensible interpretation of weak values representing an 'ordinary' property (see section 2 above for the meaning of 'ordinary' property). In particular, it casts severe doubts on whether the use of weak values really 'explains' the paradoxes – Three-Box Paradox, Hardy's Paradox – that they were supposed to explain.

**Acknowledgement** I thank S Ashhab for constructive criticism of a draft of this article

### Appendix  Relation between weak values and the Aharonov-Bergmann-Lebowitz (ABL) probability formula

In 1964 – *i.e.*, long before the concept of 'weak measurement + post-selection' was introduced – Aharonov, Bergmann and Lebowitz [23] proved an important result for ordinary, 'strong', measurement. They considered a situation very similar to the one employed in weak measurement: a system is prepared ('pre-selected') in a state $|in>$ and 'post-selected' in a state $|f>$. Then, the ABL conditional probability $prob(o_i | |f>,|in>)$ of finding a particular eigenvalue $o_i$ (assumed non-degenerate) of an operator $\hat{O}$ representing an observable $O$ in an ordinary, 'strong' measurement, intermediate in time between the pre- and the post-selection, is

$$\text{ABL:} \quad prob(o_i | |f>, in>) = \frac{|<f|o_i>|^2 \, |<o_i|in>|^2}{\sum_j |<f|o_j>|^2 \, |<o_j|in>|^2} . \quad (A1)$$

For the particular case of $\hat{O}$ being a projection/number operator, $\hat{O} = \hat{A} := |a><a|$, onto a particular eigenstate $|a>$ of an operator $\hat{\mathbb{A}}$ (the eigenvalues of $\hat{A}$ are 1 and 0, of which the latter could be degenerate even if the former is not; however, that possible degeneracy is also covered by the following treatment), the ABL formula (A1) reads

$$prob(1 | |f>, |in>) = \frac{|<f|\hat{A}|in>|^2}{|<f|\hat{A}|in>|^2 + |<f|(1-\hat{A})|in>|^2}, \quad (A2)$$

which is the probability of success in finding $|a>$, and

$$prob(0 | |f>, |in>) = \frac{|<f|(1-\hat{A})|in>|^2}{|<f|\hat{A}|in>|^2 + |<f|(1-\hat{A})|in>|^2}, \quad (A3)$$

which is the probability of failure in finding it.

In case $<f|in> \neq 0$, these expressions may be written in terms of the weak value

$$(_f\hat{A}_{in})_w = \frac{<f|\hat{A}|in>}{<f|in>} \tag{A4}$$

as

$$prob\,(1|\,|f>,|in>) = \frac{|(_f\hat{A}_{in})_w|^2}{|(_f\hat{A}_{in})_w|^2 + |[_f(1-\hat{A})_{in}]_w|^2}, \tag{A5}$$

with a corresponding expression for $prob\,(0|\,|f>,|in>)$ [3].

In particular, if $prob\,(1|\,|f>,|in>) = 1$ (or $=0$), *i.e.*, if one is certain to find (respectively not to find) the intermediate eigenstate $|a>$, one also gets the weak value $(_f\hat{A}_{in})_w = 1\,(=0)$ and *vice versa*. Furthermore, if $(_f\hat{A}_{in})_w = -1$, then the ABL probability becomes $prob\,(1|\,|f>,|in>) = ⅕$.

One may even carry the relation between the ABL probability formula and the weak value one step further: If one assumes that the imaginary part of the weak value vanishes – and this has been the case in all application of weak values to 'paradoxes' – the formula (A5) may be used to solve for $(_f\hat{A}_{in})_w$. One finds

$$(_f\hat{A}_{in})_w = \frac{\sqrt{p}}{\sqrt{p} \pm \sqrt{1-p}} \quad,\quad p := prob\,(1|\,|f>,|in>), \tag{A6}$$

where I assume $p \neq ½$ in case of the minus sign in the $\pm$-ambiguity in the denominator. I note the particular cases

$$prob\,(1|\,|f>,|in>) = 1 \Leftrightarrow (_f\hat{A}_{in})_w = 1, \tag{A7}$$

as has already been stated, and

$$prob\,(1|\,|f>,|in>) = ⅕ \Leftrightarrow (_f\hat{A}_{in})_w = -1 \text{ or } = ⅓. \tag{A8}$$

In sum, there is a very close *numerical* relation between the ABL probability $prob\,(1|\,|f>,|in>)$ of eq. (A2) applied to a number operator $\hat{A} = |a><a|$ and the weak value $(_f\hat{A}_{in})_w$: the ABL probability can be directly expresses in the corresponding weak value, and the weak value, provided it is real, can be expressed in terms of the ABL probability, albeit with a sign ambiguity. In this very restricted sense, there is a correlation between strong and weak pre- and post-selected measurements.

---

[3] These relations between weak values and ABL probabilities are by no means new; they are given, *e.g.*, in [34, 38].

Of course, *conceptually* the two notions are very different.

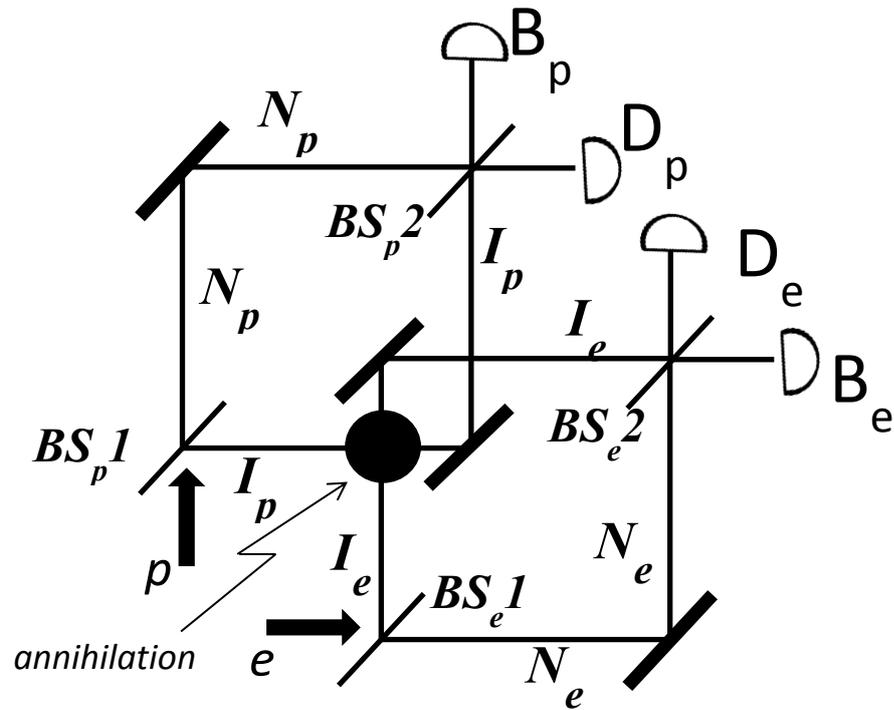

Figure 1. Schematic illustration of the experimental set-up for Hardy's Paradox. An electron (*e*) and a positron (*p*) enter each its own Mach-Zehnder interferometer with beam-splitters (*BS*), and are detected in the *B* (for 'bright') or the *D* (for 'dark') ports. They are free to move in the non-interacting arms (*N*) but annihilate each other in the intersection of the interaction arms (*I*). The paradox is that a pair appears in the *D*-ports, indicating that the particles went through the *I*-arms, even though they should then have been annihilated.

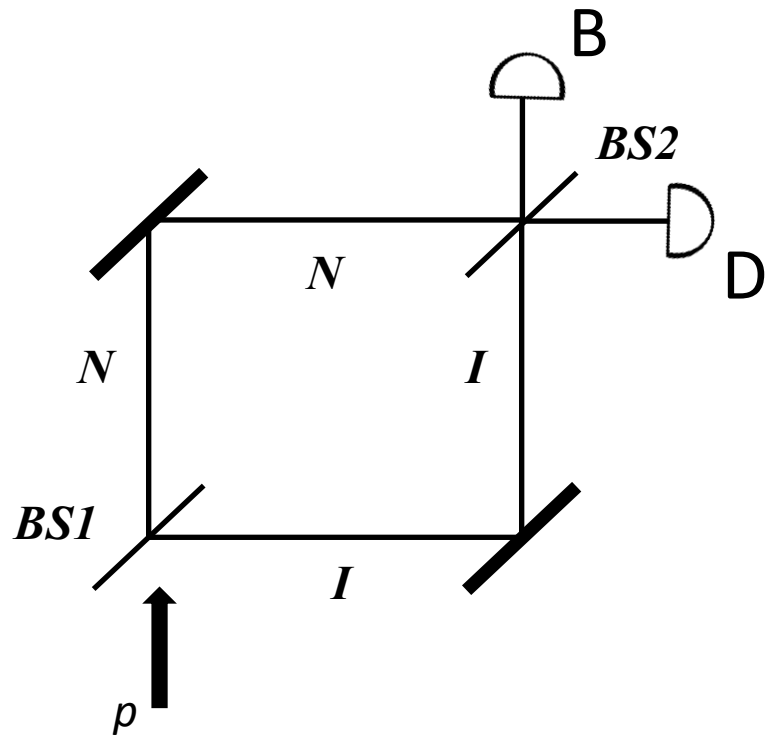

Figure 2. Schematic illustration of a single Mach-Zehnder interferometer. (See text to figure 1 for explanation of the symbols.) The beam-splitter BS1 is assumed to be a perfect, 50-50 beam-splitter, while BS2 is of a more general type (see text, in particular eq (14)).